\documentclass[reprint,amsmath,amssymb,aps,prl,nofootinbib]{revtex4-1}

\usepackage{graphicx}
\usepackage{dcolumn}
\usepackage{bm}
\usepackage{hyperref,color}

\usepackage{soul}

\def\be{\begin{equation}}
\def\ee{\end{equation}}
\def\ba{\begin{eqnarray}}
\def\ea{\end{eqnarray}}

\newcommand\nn{\nonumber}
\newcommand\q{\quad}
\newcommand{\cc}{\mathcal C}

\newcommand{\ch}{\mathcal H}

\newcommand{\cq}{\mathcal Q}
\newcommand{\calr}{\mathcal R}

\begin{document}

\title{Can chaos be observed in quantum gravity?}
\author{Bianca Dittrich$^1$\email{{bdittrich@perimeterinstitute.ca}}}
\author{Philipp A.\ H\"ohn$^2$\thanks{\texttt{p.hoehn@univie.ac.at}}}
\author{Tim A.\ Koslowski$^3$\thanks{\texttt{koslowski@nucleares.unam.mx}}}
\author{Mike I.\ Nelson$^4$\thanks{\texttt{mike@aims.edu.gh}}}
\affiliation{$^1$Perimeter Institute for Theoretical Physics, 31 Caroline Street North, Waterloo, ON N2L 2Y5, Canada}
\affiliation{$^2$Vienna Center for Quantum Science and Technology, and Institute for Quantum Optics and Quantum Information, Austrian Academy of Sciences, Boltzmanngasse 3, 1090 Vienna, Austria}
\affiliation{$^3$Instituto de Ciencias Nucleares, Universidad Nacional Aut\'onoma de M\'exico,
Apartado Postal 70-543, M\'exico D.F. 04510, M\'exico}
\affiliation{$^4$African Institute for Mathematical Sciences
P.O Box LG 197, Legon, Accra, Ghana}

\begin{abstract}
\noindent 
Full general relativity is almost certainly `chaotic'. We argue that this entails a notion of non-integrability: a generic general relativistic model, at least when coupled to cosmologically interesting matter, likely possesses neither differentiable Dirac observables nor a reduced phase space.
It follows that the standard notion of observable has to be extended to include non-differentiable or even discontinuous generalized observables. These cannot carry Poisson-algebraic structures and do not admit a standard quantization; one thus faces a {\it quantum representation problem of gravitational observables}. This has deep consequences for a quantum theory of gravity, which we investigate in a simple model for a system with Hamiltonian constraint that fails to be completely integrable. We show that basing the quantization on standard topology precludes a semiclassical limit and can even prohibit {\it any} solutions to the quantum constraints. 
Our proposed solution to this problem is to refine topology such that a complete set of Dirac observables becomes continuous. In the toy model, it turns out that a refinement to a polymer-type topology, as e.g.\ used in loop gravity, is sufficient. Basing quantization of the toy model on this finer topology, we find a complete set of quantum Dirac observables and a suitable semiclassical limit. 
This strategy is applicable to realistic candidate theories of quantum gravity and thereby suggests a solution to a long-standing problem which implies ramifications for the very concept of quantization. Our work reveals a qualitatively novel facet of chaos in physics and opens up a new avenue of research on chaos in gravity which hints at deep insights into the structure of quantum gravity.
\end{abstract}

\maketitle

\noindent The canonical description of gauge theories, classical and quantum alike, with totally constrained Hamiltonian encodes the dynamics of the system in `constants of motion' \cite{Dirac,Henneaux:1992ig}, so-called Dirac observables. One interprets a complete set of Dirac observables as all that can objectively be predicted about the classical or quantum system. Much has been written about Dirac observables for general relativity \cite{Kuchar:1991qf,Isham:1992ms,Anderson:2012vk,relrov,Rovelli:1989jn,Rovelli:1990jm,Rovelli:1990ph,Rovelli:1990pi,Rovelli:2004tv,Dittrich:2004cb,Dittrich:2005kc,Dittrich:2006ee,Dittrich:2007jx,Bojowald:2010xp,Bojowald:2010qw,Hohn:2011us,Hajicek:1995en,Hajicek:1996xk,Tambornino:2011vg,Thiemann:2007zz,Giddings:2005id,Marolf:1994nz,Marolf:2009wp,Hajicek:1994py,Anderson:2013wua}, which involves the implementation of invariance under spacetime diffeomorphisms. 

It has however often been overlooked that Dirac observables may not always exist as differentiable phase space functions. This occurs in analogy to classical chaotic systems \cite{arnold2007mathematical,ottchaos,gutzwillerbook,berrychaos} when the flow generated by the Hamiltonian constraint is sufficiently complicated \cite{Hohn:2011us,Hajicek:1995en,Hajicek:1996xk,Dittrich:2015vfa,UnruhWald,Kuchar:1993ne,Torre:1993jm,Anderson:1995tu,Smolin:2000vs}. 
Specifically, there are strong hints that full general relativity is non-integrable or even chaotic \cite{Torre:1993jm,Anderson:1994eg,Barrow:1981sx,Cornish:1997ah,Page:1984qt,Kamenshchik:1998ue,Hohn:2011us,Misner:1969hg,ringstrom,Chernoff:1983zz,Cornish:1996yg,Cornish:1996hx,Motter:2000bg,Belinsky:1970ew,Barrow:1997sb}, and that a generic general relativistic model with cosmologically interesting matter is likely to admit neither differentiable Dirac observables nor a symplectic reduction. While we shall discuss evidence for the latter below, we refer the reader to \cite{Dittrich:2015vfa} for a more in-depth discussion. Tellingly, differentiable Dirac observables for full general relativity are not known \cite{Tambornino:2011vg,Anderson:2013wua} due to the quadratic nature of the Hamiltonian constraint \cite{Arnowitt:1962hi,Thiemann:2007zz} apart from boundary charges (see e.g.\ \cite{Arnowitt:1962hi} for asymptotically flat and \cite{Henneaux:1984xu,Henneaux:1985tv,Anderson:1996sc} for asymptotically Anti-deSitter) or in dust filled spacetimes \cite{Brown:1994py,Husain:2011tk}. This is deeply intertwined with the absence of good (monotonic) time or clock functions or, equivalently, good gauge fixing conditions from a generic general relativistic model \cite{Kuchar:1991qf,Isham:1992ms,Bojowald:2010xp,Bojowald:2010qw,Hohn:2011us,Hajicek:1995en,Hajicek:1996xk,Giddings:2005id,Dittrich:2015vfa}.

But if differentiable Dirac observables are absent, what is then observable? What is the physical interpretation of such putative observables? And what are the consequences for a quantum theory? 

Given that chaotic gravitational models turn out to be analytically too intricate, we address these crucial questions in the probably simplest non-trivial toy model which, however, qualitatively mimics dynamical properties of a chaotic cosmological model. The employed model is the reparametrization-invariant description of two free particles on a circle with fixed energy, in which the angular momentum-like Dirac observable is discontinuous in the standard topology. We show that a quantization using standard techniques precludes a semiclassical limit or even any solutions to the quantum constraints. This confirms heuristic worries in the older quantum gravity literature \cite{UnruhWald,Kuchar:1993ne,Torre:1993jm,Anderson:1995tu,Smolin:2000vs} and mimics the breakdown of semiclassical wavepackets for the Wheeler-DeWitt equation of the chaotic Friedmann-Robertson-Walker model with a massive scalar field reported in \cite{Hohn:2011us,Hawking,Kiefer:1988tr,Cornish:1997ah} (see also references therein). However, we also identify the root of the problem: the topology underlying the quantization. If chosen unsuitably, it may not admit sufficiently many solutions to the quantum constraints  in order to allow for semiclassical states. This aspect has been neglected in older heuristic discussions \cite{UnruhWald,Kuchar:1993ne,Torre:1993jm,Anderson:1995tu,Smolin:2000vs} and leads to our proposal for a resolution: to adapt the method of quantization to the (dis-)continuity of the observables which one wishes to represent in the quantum theory. We show that, basing the quantization on a polymer-like topology (see e.g.\ \cite{Hossain:2010wy,Laddha:2010hp,Corichi:2007tf}), similar to the one used in loop quantum cosmology \cite{Bojowald:2008zzb,Ashtekar:2011ni,Ashtekar:2003hd} and a novel representation of loop quantum gravity \cite{Dittrich:2014wpa,Bahr:2015bra}, resolves the continuity problem and the topology refined quantization admits the expected semiclassical limit. This leads to the main conclusion of the paper: {\it The quantization of systems with Hamiltonian constraints needs to be based on a topology that is fine enough to allow for a complete set of continuous gauge invariant observables.}

Our results suggest profound repercussions for non-integrable constrained systems and, given the evidence of non-integrability in general relativity, thereby also for the quantization of gravity.

\section{(Non-)integrability and observables}

A rigorous definition of chaotic constrained dynamics is not relevant for our purposes. However, the statement of the problem and our recipe require the (more general) notion of weak non-integrability and a precise link between topology and quantization.
\begin{center}

{\it Weak (Non-) Integrability:}

\end{center}
\noindent The important obstruction to quantization that we explore in this letter is due to the absence of complete integrabilty of the Hamiltonian constrained system. 

 Consider a (finite-dimensional) system on a $2N$-dimensional kinematical phase space $\mathcal P$ subject to $m_1$ first and $2\,m_2$ second class irreducible constraints. A `Dirac observable' is a function defined on the constraint surface $\mathcal C \subset \mathcal P$ which Poisson-commutes with the first class constraints and is differentiable (such that the Poisson brackets between observables exist). In analogy to unconstrained, autonomous Hamiltonian systems \cite{arnold2007mathematical,ottchaos,gutzwillerbook,berrychaos}, we shall say that this system is {\bf{weakly integrable}} \cite{Dittrich:2015vfa} if there exist $2(N-m_1-m_2)$ independent Dirac observables $O_i$ which are also independent of the constraints, and if $N-m_1-m_2$ of these $O_i$ are weakly in involution $\{O_i,O_j\}\approx0$, where $\approx$ is a weak equality (i.e., equality on $\cc$).

This ensures that a weakly integrable system is reducible, i.e.\ there exists a symplectic reduced phase space (locally given by $\mathcal P_R:=\mathcal C/\gamma$ where $\gamma$ denotes the gauge orbit \cite{Henneaux:1992ig,Dittrich:2015vfa,Hajicek:1995en,Hajicek:1996xk}). Said simply: There exist a maximal independent set of Dirac observables which provides good coordinates for the reduced phase space and there is an induced Poisson structure between these.

By contrast, a system that fails to be weakly integrable will in general not possess a symplectic reduced phase space. There is not a sufficient number of differentiable Dirac observables to provide coordinates on the space of solutions which consequently  does not possess an induced Poisson structure (see e.g.\ \cite{Henneaux:1992ig,Dittrich:2015vfa,Hajicek:1995en,Hajicek:1996xk}). 

Nevertheless, the absence of Dirac observables does {\it not} render systems that fail to be weakly integrable devoid of gauge invariant observables: functions that are constant along gauge orbits {\it always} exist. However, they may either be non-differentiable (see e.g.\ \cite{Dittrich:2015vfa,UnruhWald,Hajicek:1995en,Hajicek:1996xk}) or only defined locally on $\mathcal C$ (a generalized Darboux theorem ensures their existence \cite{Tyutinbook,Dittrich:2015vfa}). This has deep consequences especially for the quantum theory: a globally defined Poisson algebra of gauge invariant observables does {\it not} exist in weakly non-integrable systems. These are consequences of {\it global} constraint surface properties -- and quantization strongly depends on them \cite{isham2,Dittrich:2007th}.

We emphasize that there is a crucial distinction between this notion of {\it weak} non-integrability and non-integrability on a reduced (or more generally unconstrained) phase space \cite{Dittrich:2015vfa}. Technically, the latter actually requires the system to be weakly integrable such that a reduced phase space exists with a non-integrable unconstrained dynamics on it, while the former implies the absence of a symplectic reduced phase space altogether.\footnote{E.g., consider an unconstrained non-integrable system subject to
a (time independent) Hamiltonian $H$. Parametrizing this system would yield a Hamiltonian constraint $C=p_t+H$, where $p_t$ is the momentum conjugate to the (now dynamical) time variable $t$. The resulting constrained system is evidently weakly integrable because $t=const$ yields a global gauge fixing and global Dirac bracket, yet has a non-integrable dynamics generated by $H$ on the reduced phase space.}

But there is also a conceptual difference. Firstly, an absence of differentiable constants
of motion on an unconstrained phase space does not pose a conceptual challenge because one does not need to solve the (reduced) dynamics in order to access the physical degrees of freedom. By contrast, in a totally constrained system, one firstly needs to solve the dynamics in order to access
the physical degrees of freedom that are required for interpreting the dynamics. Thus, weak non-integrability and the absence of differentiable Dirac observables have more severe repercussions that can turn into a quantum representation problem of the surviving non-differentiable observables.

\begin{center}
{\it Topology and Quantization:}
\end{center}
Canonical quantization requires a polarization, which usually amounts to choosing a differentiable manifold $\mathcal Q$ as configuration space within phase space $\mathcal P$. The Poisson-action of the momenta then defines vector fields on $\mathcal Q$ which can be integrated to a group action $G$ on $\mathcal Q$. The topology of $\mathcal P$ induces topologies on $\mathcal Q, G$. Quantization assigns to each continuous complex valued function $f\in C(\mathcal Q)$ (or a dense subset thereof) and to each element $g\in G$ an operator in the quantum algebra $\mathfrak A$, respectively. $\mathfrak A$ is represented on a Hilbert space $\mathcal H$  which equips $\mathfrak A$ with the strong operator topology. Topology enters quantization by demanding that the quantization map  $C(\mathcal Q) \to \mathfrak A$ and $ G \to \mathfrak A$ is continuous. This continuity requirement  ensures that the position and momentum operators are densely defined.

In this letter, we use the expression `the topology that underlies a quantization', the precise meaning of which is defined by the requirement that the quantization map is continuous.

\begin{center}
{\it Weak Non-Integrability in General Relativity:}
\end{center}

We will now explain why our notion of weak integrability is not satisfied by general relativity with phenomenologically interesting matter content, despite the existence of theorems (e.g., see \cite{fishermarsden,isenberg,andersson}) that show that the reduced phase space of general relativity may exist for various types of conformal matter. In fact, we argue that general relativity fails to be weakly integrable when coupled to any matter content that produces realistic cosmological predictions. Our argument proceeds in several steps:

The phase space reduction discussed in \cite{fishermarsden,isenberg,andersson} relies on the existence and uniqueness of the solution of the Hamilton constraint of general relativity on compact slices in terms of the conformal method. Technically, this means that York's modification of the Lichnerowicz equation possesses a unique positive solution \cite{YorkNiall} for the conformal factor, which in turn implies that constant-mean-curvature (CMC) slicing defines good gauge conditions and that York time may be a good temporally global clock variable. Effectively, the results of \cite{fishermarsden,isenberg,andersson} thereby rely on the existence of a good time function and associated to this on good CMC-gauges. However, this requires that the matter Hamiltonian possesses a specific transformation behavior under conformal transformation, namely it necessitates conformal matter. This behavior precludes, among other restrictions, non-conformal matter and thus, e.g., a cosmological constant, scalar field potentials, mass terms and Yukawa interactions, which in particular precludes matter content that can drive inflation \cite{Dittrich:2015vfa}. Indeed, for non-conformal matter on compact slices, the York-Lichnerowicz equation possesses generally non-unique solutions so that York time may not be a good clock (example below) \cite{isenberg2}.


With this part of the argument we have established that it is not necessarily true that general relativity possesses a symplectic reduced phase space because crucial conditions going into the theorems of \cite{fishermarsden,isenberg,andersson} are generically violated in solutions with cosmologically interesting matter. In the following we need to establish that there are cases in which symplectic reduction indeed does not exist when cosmologically interesting matter is present. For example, the closed Friedmann-Roberson-Walker (FRW) cosmology, when minimally coupled to a homogeneous massive scalar field, has been shown to be chaotic \cite{Page:1984qt,Cornish:1997ah,Kamenshchik:1998ue}. A generic solution features both an initial and final singularity, but the model admits an infinite number (but measure zero set) of solutions which bounce perpetually among finite extrema in both the scale factor $a$ and the scalar field $\phi$, of which a countable subset is periodic and an uncountable subset is aperiodic.\footnote{However, we note that also a generic singular trajectory behaves `chaotically' in the sense that it will bounce and oscillate a large number of times in between various extrema before running into a singularity \cite{Page:1984qt,Cornish:1997ah,Kamenshchik:1998ue}.} This leads to a fractal structure in the space of solutions \cite{Page:1984qt,Cornish:1997ah,Kamenshchik:1998ue}. This constitutes an example of a general relativistic model on compact slices with non-conformal matter that explicitly violates conditions underlying the theorems in \cite{fishermarsden,isenberg,andersson}. Indeed, York time $\tau_{York}$ in this FRW model is proportional to ${\dot{a}}/{a}$ and due to the bounces $\tau_{York}=0$ is crossed uncountably many times for non-singular solutions. CMC-slicing thus does not provide good gauges in this model.

A generic perpetually bouncing trajectory remains in a compact region of the two-dimensional non-compact configuration space \cite{Page:1984qt,Kamenshchik:1998ue}. While aperiodicity and compact dynamics do not in general imply ergodicity of the trajectory in its corresponding compact region, it is highly plausible that they do in this cosmological model because the aperiodic trajectories do not admit cycles and instead bounce effectively like a billiard \cite{Page:1984qt,Cornish:1997ah,Kamenshchik:1998ue}.

Now, a single ergodic trajectory $\gamma$ in a compact region $\mathcal R_\gamma\subset \mathcal Q$ of configuration space is sufficient to imply discontinuity of any non-trivial configuration observable $O$ in this region  which is constant along the trajectories but not constant on $\cq$. To see this, restrict $O$ to $\calr_\gamma$, $O_{\calr_\gamma}:\calr_\gamma\rightarrow \mathbb{R}$, and let $O(\gamma)$ be the value of $O$ on $\gamma$. Then there does not exist an open neighborhood in $O_{\calr_\gamma}(\calr_\gamma)\subset\mathbb{R}$ which does {\it not} contain $O(\gamma)$ and which also has an open neighborhood $U\subset\calr_\gamma$ as pre-image because any such $U\subset\calr_\gamma$ will be intersected by $\gamma$. Accordingly, $O_{\calr_\gamma}$ must be discontinuous on $\calr_\gamma$.

To conclude the argument for weak non-integrability, we need to consider general phase space observables. The analyses in \cite{Page:1984qt,Cornish:1997ah,Kamenshchik:1998ue} also suggest that the momenta remain bounded for generic aperiodic, perpetually bouncing solutions such that these also remain restricted to compact subregions of the constraint surface. Repeating then the previous argument indicates strongly that also any phase space observable must be discontinuous on compact subregions of $\cc$. This is compelling (yet not exhaustive) evidence that this closed FRW model with a homogeneous massive scalar field is weakly non-integrable.

Next we argue that a weakly non-integrable cosmological model also implies weak non-integrability of the {\it full} theory (for analogous matter content). Suppose we had a weakly integrable system, so there exists a set of differentiable Dirac observables that separate the gauge-orbits on the constraint surface. Let us now consider a subset of the constraint surface that itself is foliated into gauge orbits. Then it follows from the existence of sufficiently many orbit separating, differentiable Dirac observables of the full system that their restriction to the subset again separates orbits of this subset. Thus, if the restriction map from the full constraint surface to the subset is differentiable, one obtains that the subsystem is weakly integrable as well. Conversely, if an (in this sense) differentiably embedded subsystem is not weakly integrable then the full system is not weakly integrable. This suggests that full general relativity with cosmologically interesting matter content is not weakly integrable, because it contains a subsystem that has an effective billiard dynamics. 

Since the dynamics of chaotic general relativistic models are analytically not fully tractable, it is necessary to employ a toy model to study the classical and quantum consequences of weak non-integrability explicitly. Our example will emulate key features of the above FRW modelÕs dynamics: on a compact configuration space we will have both periodic as well as (uncountably many) aperiodic solutions. The aperiodic solutions will be ergodic on configuration space, precluding a smooth reduced phase space to exist. Our first aim is to explore the consequences for the quantization of such a constrained system.

\section{Toy Model}

\begin{figure}[hbt!]
\begin{center}
{\includegraphics[scale=1]{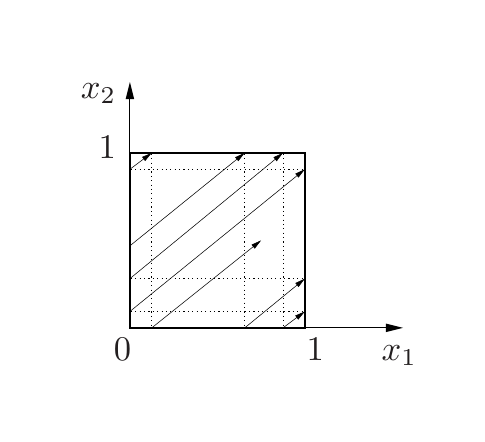}}
\caption{\small A trajectory on the torus with angle given by $\gamma$. $n_i$ increases by $1$ whenever $x_i=1$ is reached.}\label{fig_1}
\end{center}
\end{figure}

We consider two free point particles of masses $m_i$ on a unit circle and require that the total energy of the system is fixed to $E>0$ which defines the Hamiltonian constraint (or the classical time-reparametrization generator)
\begin{equation}
 C=\frac{p_1^2}{2 m_1}+\frac{p_2^2}{2m_2}-E\approx 0.\label{con}
\end{equation}
The ensuing constraint surface is compact $\mathcal C=\mathcal S^1 \times \mathcal T^2$, where the torus is parametrized by the configurations $(x_1,x_2)$ and $\mathcal S^1$ by the angle $\phi:=\textrm{atan2}(m_2 p_1,m_1 p_2)$ of the trajectory {(where $\textrm{atan2}$ is the two-argument arc-tangent)}. Note that the magnitude {$|p_i|$} is determined by $C$. The classical dynamics is
\begin{equation}
 x_i(t)=\frac{p_i}{m_i}t+x_{i0}-n_i,
\end{equation}
where $n_i=\lfloor \frac{p_i}{m_i}t+x_{i0} \rfloor$ denotes the winding number in $x_i$ (see figure \ref{fig_1}). All orbits with $\tan \phi \in \mathbb Q$ are closed and periodic, so-called {\it resonant tori}, while each trajectory with $\tan \phi \not\in \mathbb Q$ fills the torus densely, the so-called {\it non-resonant tori}.

The angle $\phi$ is a Dirac observable. A second independent phase space function that is constant along the orbits is $M=(x_1+n_1)p_2/m_2-(x_2+n_2)p_1/m_1$, but it is not continuous since the winding numbers $n_i$ are discontinuous phase space functions.  There cannot exist any continuous gauge invariant observable with configurational dependence because for $\tan\phi \not\in \mathbb Q$ the trajectories fill the torus densely, such that any continuous gauge invariant observable needs to be independent of the positions \cite{Dittrich:2015vfa}. Moreover, the space of solutions is neither a manifold nor Hausdorff \cite{Dittrich:2015vfa}, such that Marsden-Weinstein reduction fails. The system fails weak integrability by admitting only a single (independent) momentum Dirac observable. 
Gravity seems to behave worse since there is strong evidence that the space of solutions of the chaotic closed FRW model with minimally coupled massive scalar field features a fractal structure \cite{Page:1984qt,Cornish:1997ah,Kamenshchik:1998ue}.

\section{Quantization (Standard Topology)}
 A reduced quantization of the space of solutions is impossible because it is not a phase space.
But standard Dirac quantization of the toy model is possible and based on choosing $\mathcal Q=\mathcal T^2$ to be the configuration space and $G$ to be the group of translations on $\mathcal Q$. This quantization is represented on the kinematical Hilbert space $\mathcal H_{kin}=L^2(\mathcal Q)$;  the topology ensures that the momentum operators $\hat p_i\,\psi=-i\hbar\partial_i\,\psi$ are densely defined on the span of the momentum eigenstates
\begin{equation}
 \psi_{k_1,k_2}(x_1,x_2)=\exp(2\pi\,i(k_1 x_1+k_2 x_2)), \q\q k_i\in\mathbb{Z}
\end{equation}
which  diagonalize the constraint operator
\begin{equation}
 \hat C = \frac{\hat p_1^2}{2 m_1}+\frac{\hat p_2^2}{2m_2}-E\,\hat{\mathbb I},
\end{equation}
which has a {\it discrete} spectrum
\begin{equation}
 \Delta(\hat C)=\left\{(2\pi\hbar)^2\left(\frac{k_1^2}{2m_1}+\frac{k_2^2}{2m_2}\right)-E:k_i\in \mathbb Z\right\}.
\end{equation}
 The constraint kernel has dimension $0,1,2$ or $4$, depending on the value of $\epsilon:=\frac{m_1\,E}{2(\pi\hbar)^2}$ whenever $\gamma \not\in\mathbb Q$, where $\gamma=\frac{m_2}{m_1}$. For $\gamma\in \mathbb Q$ the problem of finding the dimension of the constraint kernel results in a rich Diophantine problem. In this case, the constraint kernel can be larger, but it turns out (due to Jacobi \cite{Jacobi1829:aa,Dittrich:2015vfa}) that  its dimension is bounded by $\log\epsilon$, such that the kernel will be very small even for macroscopic $\epsilon$. (For comparison, the degeneracy of energy eigenstates of two harmonic oscillators scales with $\epsilon$.)

The kernel is generically too small for admitting semiclassical wavepackets  and there is no conjugate pair of Dirac observables to peak on. Moreover, dynamical coherent states, which have to be wave packets of the form $\psi(x_1,x_2)=\sum_k a_k e^{in_k(x_2+\alpha x_1)}$ require a linear dispersion relation which is prohibited by the Hamiltonian constraint $\hat C$ \cite{Dittrich:2015vfa}.  Hence, one cannot build dynamical coherent states even when $\gamma\in \mathbb Q$ and $\epsilon$ is macroscopic. 

More extremely, modifying the classical dynamics by setting $E=0$ and flipping the relative sign in (\ref{con}) yields a quantum constraint $k_2=\pm\sqrt{\gamma}\,k_1$ which has {\it no} solutions for $\sqrt{\gamma}\notin\mathbb{Q}$. (Even small quantum corrections to the energy would not generate more than four solutions.)

\section{Quantization (Refined Topology)}

In the previous section we found that the scarcity of smooth Dirac observables leads to a physical Hilbert space that is too small for a satisfactory semiclassical limit. This suggests to consider a discrete topology on the configuration space (i.e.\ the torus), such that  observables, discontinuous with respect to the standard topology, become continuous -- and representable in the quantum theory. 

We consider a {kinematical} Hilbert space of periodic functions $\psi(x_1+1,x_2)=\psi(x_1,x_2+1)=\psi(x_1,x_2)$ with orthonormal basis
$\psi_{\alpha_1,\alpha_2} (x_1,x_2) = \delta_{\alpha_1,x_1} \delta_{\alpha_2,x_2}$ 
and Kronecker delta  functions 
 \ba\label{d4}
 \delta_{\alpha,\alpha'}    &=&
 \begin{cases}
 1 \q \text{for}\, \alpha=\alpha' \q \text{mod} \q 1 \\
 0\q \text{otherwise.}
 \end{cases}
 \ea

Functions in this Hilbert space are vanishing almost everywhere and discontinuous with respect to the Lebesgue measure. Thus, momenta as derivative operators  are not well-defined, in contrast to translation operators  
%
 %
%
 \ba
R_1^\mu \psi (x_1,x_2) = \psi(x_1+\mu,x_2),\\\nn
R_2^\mu \psi  (x_1,x_2) = \psi(x_1,x_2+\mu)    .
\ea
This introduces a novel (length) parameter $\mu$ whose interpretation comes later. The square of the momentum $\frac{\hat{p}^2_a}{2}$ is quantized as
$
S^\mu_a :=   \frac{-\hbar^2}{ 2 \mu^2}   \left( R_a^{+\mu}  + R_a^{-\mu} - 2 \mathbb{I} \right)  ,
$
 yielding the constraint (for this section with $m_1=m_2=1$) 
\ba\label{constraintD}
C^\mu &=& S^\mu_1+ S^\mu_2 - E.
\ea

Let us discuss the spectrum of this constraint and, more generally, the translation operators 
for $\mu\notin\mathbb{Q}$ (for $\mu\in\mathbb{Q}$, see \cite{Dittrich:2015vfa}). Consider periodic functions in only one variable $\psi(x)$. The Hilbert space splits into invariant subspaces ${\cal H}^\mu_\alpha$
\ba\label{d19}
{\cal H}^\mu_\alpha=\text{clos}(\text{span}\{ \psi_{\alpha-N \mu } \,|\, N \in \mathbb{Z}\} ) .
\ea
(The labels are to be understood as $\text{mod}\, 1$.)  The $\alpha$ label the invariant subspaces and denote  equivalence classes: $\alpha'\sim\alpha$ if $\alpha=\alpha'+ N \mu \, \text{mod}\, 1$ for some $N \in \mathbb{Z}$. 

The spectrum of $R^\mu$ on each of the ${\cal H}^\mu_\alpha$ is  continuous and given by $U(1)$ \cite{Bahr:2015bra}. 
This can be understood by considering the translation operator $\exp(i \mu \hat p)$ in the usual quantization (on $S^1$): the spectrum ${\mathbb Z}$ of $\hat p$ winds around the unit circle and due to  $\mu\notin \mathbb{Q}$ fills it densely. In the standard quantization the spectrum would be classified as discrete as the eigenvectors are normalizable. Here, by contrast, the spectrum is continuous due to generalized eigenvectors:
\ba \label{d20}
u_{\alpha,\rho}(x) \,=\, \sum_{ l \in \mathbb{Z}}  e^{2\pi i  \, l \rho}    \delta_{ \alpha + l\mu , x},\q\q\rho\in [0,1).
\ea

The constraint (\ref{constraintD}) combines the (commuting) operators $R^{\pm \mu/2}_{a}$ and has a {\it continuous} spectrum 
\ba
\frac{\hbar^2}{ 2 \mu^2} \left(4 - 2 \cos(2\pi \rho_1) -2\cos(2\pi \rho_2 ) \right)  - E,\;\rho_i \in [0,1).\q\;\;
\ea
We can restrict to one choice of $(\alpha_1,\alpha_2)$ which yields a superselection sector $\ch^\mu_{\alpha_1}\otimes\ch^\mu_{\alpha_2}$ with respect to the momentum operators and thus a separable Hilbert space. 

The spectrum is contained in $[-E, \frac{ 8 \hbar^2}{ 2 \mu^2}  -E ]$. Hence $\mu$ determines the maximal energy and, by duality, the minimal step-size. This is analogous to  loop quantum cosmology, where the minimal step-size is set by the Planck scale and  the energy density is bounded by the Planck density \cite{Bojowald:2008zzb,Ashtekar:2011ni,Ashtekar:2003hd}.

After fixing one superselection sector $(\alpha_1,\alpha_2)$, the solutions to $C^\mu\psi=0$ for a fixed (allowed) $E$ are labelled by  the energy $e_1$ for the $x_1$ particle with $\text{max}(0,E-\frac{ 4 \hbar^2}{ 2 \mu^2}) \leq e_1 \leq E$. (There exist  sign degeneracies.)
Crucially, there are infinitely many solutions. 
But these are non--normalizable with respect to the kinematical inner product such that we need a physical inner product in which they are normalizable. 

This can be constructed using the $\rho$-representation, given by the spectral decomposition of the translation operators on a fixed ${\cal H}^\mu_{\alpha_1} \otimes {\cal H}^\mu_{\alpha_2}$
\ba\label{rhorep}
\psi \,=\, \int_{[0,1) \times [0,1) }\!\!\!\!\!\!\!\!\! {\rm d} \rho_1  {\rm d} \rho_2\,\,    u_{\alpha_1, \rho_1}  u_{\alpha_2, \rho_2}\,\,   \langle u_{\alpha_1,\rho_1}u_{\alpha_2,\rho_2} \,|\,\psi \rangle. \q 
\ea
The method combines refined algebraic \cite{Ashtekar:1994kv,Louko:1999tj,Marolf:1995cn,Marolf:2000iq} and  master constraint \cite{Dittrich:2004bn,Dittrich:2004bq} quantization and results in a physical Hilbert space that can be expressed as a $L^2$ Hilbert space over a continuous (momentum) parameter $\rho$ \cite{Dittrich:2015vfa}.

Using the $(\rho_1,\rho_2)$-representation defined in (\ref{rhorep}), one finds that the following observable
\ba\label{defqM}
\hat M\,:=\,
\frac{i}{2\pi}\left(
 \sin(2\pi\rho_2) \frac{\partial}{ \partial \rho_1} - \sin(2\pi\rho_1) \frac{\partial}{ \partial \rho_2} 
\right)
\ea
commutes with the constraint and hence provides a quantum Dirac observable, which is analogous to the classical discontinuous observable $M$, which does not admit a quantization in the standard topology. 

The quantization based on the discrete topology features superselection sectors (for the translation operators) labelled by $(\alpha_1,\alpha_2)$. States in ${\cal H}^\mu_{\alpha_1} \otimes {\cal H}^\mu_{\alpha_2}$ are restricted to a lattice with lattice constant $\mu$ and one lattice vertex in $(\alpha_1,\alpha_2)$. Thanks to $\mu\notin\mathbb{Q}$, this lattice fills the torus densely. Nevertheless, one can `unwind' the periodic variable $x_i$ by replacing it with $l_i$ defined through  $x_i=(\alpha_i + l_i \mu)\, \text{mod}\, 1$.  The $l_i \in {\mathbb Z}$ parametrize the vertices in the lattice one-to-one and generate a $l_i$--representation (by Fourier transforming the $(\rho_1,\rho_2)$-representation) in which states need no longer be periodic. The crucial point is that the winding numbers $n_i$ are quite regular functions of the $l_i$. This allows us to represent $M$ in this quantization. 

In summary, the physical Hilbert space is infinite-dimensional and enables us to represent both a momentum and a configurational Dirac observable. The result can be viewed either as a quantization of a classically modified dynamics (which is still weakly non--integrable) or as a discrete quantization of the original unmodified system. Indeed, choosing the parameter $\mu$ sufficiently small and for momenta considerably smaller than the maximal energy scale set by $\mu$, the two systems possess approximately the same dynamics.

\section{Transition Amplitudes and Classicality}

A useful tool in the exploration of a semiclassical behavior is the physical transition amplitude $W(\vec x_1,\vec p_1;\vec x_2,\vec p_2)$ between kinematic states
\begin{equation}\label{equ:W}
 W=\frac{\langle(\vec x_1,\vec p_1),\hat P (\vec x_2,\vec p_2)\rangle}{\sqrt{\langle(\vec x_1,\vec p_1),\hat P (\vec x_1,\vec p_1)\rangle\langle(\vec x_2,\vec p_2),\hat P (\vec x_2,\vec p_2)\rangle}}, 
\end{equation}
where $|(\vec x,\vec p)\rangle$ denote kinematical coherent states that are peaked at a phase space point $(\vec x,\vec p)$ and $\hat P$ denotes the projector to the kernel of the Hamiltonian constraint. A quantum system admits a good semiclassical limit of the Hamiltonian constrained system if $W(\vec x_1,\vec p_1;\vec x_2,\vec p_2)\approx 1$ when $(\vec x_1,\vec p_1)$ and $(\vec x_2,\vec p_2)$ lie on the same trajectory and decreases rapidly when one moves one of $(\vec{x}_i,\vec{p}_i)$ away from the trajectory. 

For the standard quantization one can show that $W\approx1$ only when the momenta $\vec p_1$ and $\vec p_2$ are simultaneously equal to one of the solution momenta $\vec k \in \mathbb Z^2$ to the quantum constraint. This precludes a sufficiently peaked position dependence \cite{Dittrich:2015vfa} and is another way of seeing that the standard quantization does not admit a semiclassical limit.  

The situation differs when one bases quantization on the refined topology. The ingredients of the definition (\ref{equ:W}) of the transition amplitude can also be constructed for the quantization based on the refined topology and it turns out that $W$ attains a good semiclassical limit in this case \cite{Dittrich:2015vfa}: The physical transition amplitude between kinematical semi-classical states is maximal if the kinematical semiclassical sates are peaked on points in the same orbit and decay rapidly as one moves one of the points away from the orbit on which the other point lies.

\section{Conclusions}

Quantizing weakly non-integrable systems with standard methods ignores the discontinuity of their observables, yields a small or even empty physical Hilbert space and no semiclassical limit. This puts the breakdown of semiclassical states reported in \cite{Hohn:2011us,Hawking,Kiefer:1988tr,Cornish:1997ah} for the chaotic FRW model with scalar field into a novel perspective.
These troubles are a direct consequence of the scarcity of Dirac observables and differ qualitatively from the  
situation in {\it unconstrained} chaotic systems where 
the quantum representation problem of observables does {\it not} arise and one generally obtains
at least a short time coherence of wave packets \cite{ottchaos,gutzwillerbook,berrychaos}. This reveals a previously unknown aspect of chaos (or non-integrability) in physics: a quantum representation problem of observables in consequence of diffeomorphism symmetry (here reparametrization invariance) which thus is special to gravitational systems.
Our findings indicate that, {\it for weakly non-integrable systems, one needs to adapt the very method of quantization to the discontinuous observables in order to obtain a functional quantum theory with interesting observables and semciclassical limit.} Promisingly, the general strategy of topology refinement is not specific to our toy model but also applicable to realistic candidate theories for quantum gravity. 

 The alternative quantization method  used in our investigations is that of `polymer quantization' \cite{Hossain:2010wy,Laddha:2010hp,Corichi:2007tf} 
which is employed in loop quantum cosmology and gravity \cite{Bojowald:2008zzb,Ashtekar:2011ni,Ashtekar:2003hd,Bahr:2015bra}. However,  its remedies come at a physical price: the spectra of operators become `Bohr compactified', superselection sectors arise and one has to interpret the additional quantization parameter $\mu$. Promisingly, the minimal `step size' $\mu$ can be related to the Planck scale in a gravitational context \cite{Bojowald:2008zzb,Ashtekar:2011ni,Ashtekar:2003hd}.

Since full general relativity is {almost certainly `chaotic' \cite{Dittrich:2015vfa,Torre:1993jm,Anderson:1994eg,Barrow:1981sx,Cornish:1997ah,Page:1984qt,Kamenshchik:1998ue,Hohn:2011us,Misner:1969hg,ringstrom,Chernoff:1983zz,Cornish:1996yg,Cornish:1996hx,Motter:2000bg,Belinsky:1970ew,Barrow:1997sb} and, owing to our arguments, likely features weak non-integrability for cosmologically interesting matter}, these observations imply deep ramifications for a quantum theory of gravity. These investigations 
thus warrant an extension to the field theory case and, ultimately, general relativity. Also, the consequences of weak non-integrability for a path integral quantization, which should produce a projector onto the solutions of the quantum constraints \cite{Halliwell:1990qr,Rovelli:1998dx}, remain to be clarified (see also~\cite{Amaral:2016hud}). Finally, how far does one need to refine the topology underlying the constraint quantization in order to obtain a functional and interesting quantum theory?

\section*{Acknowledgements}

We thank Benjamin Bahr, Ted Jacobson, Aldo Riello, Carlo Rovelli and Bill Unruh for discussion. Research at Perimeter Institute is supported by the Government of Canada through Industry Canada and by the Province of Ontario through the Ministry of Research and Innovation. The project leading to this publication has received funding from the European Union's Horizon 2020 research
and innovation programme under the Marie Sklodowska-Curie grant agreement No 657661 (awarded to PH). MN is grateful
to AIMS Ghana for a stipend and was also supported by an NSERC grant awarded to BD. TK received financial support from the
Foundational Questions Institute (fqxi.org) and thanks the Perimeter Institute for its hospitality during the start of this work.

\end{document}